# The Space Optical Clocks Project:
## Development of high-performance transportable and breadboard optical clocks and advanced subsystems

S. Schiller, A. Görlitz, A. Nevsky, S. Alighanbari, S. Vasilyev,
C. Abou-Jaoudeh, G. Mura, T. Franzen,
Heinrich-Heine-Universität Düsseldorf (D);
U. Sterr, S. Falke, Ch. Lisdat,
Physikalisch-Technische Bundesanstalt (D);
E. Rasel, A. Kulosa,
Leibniz Universität Hannover (D);
S. Bize, J. Lodewyck,
Observatoire de Paris (F);
G. M. Tino, N. Poli, M. Schioppo,
Università di Firenze and LENS (I);
K. Bongs, Y. Singh,
University of Birmingham (UK);
P. Gill, G. Barwood, Y. Ovchinnikov,
National Physical Laboratory Teddington (UK);

J. Stuhler, W. Kaenders,
TOPTICA Photonics AG (D);
C. Braxmaier,
EADS Astrium Friedrichshafen (D);
R. Holzwarth,
Menlo Systems GmbH (D);
A. Donati,
Kayser Italia Srl (I);
S. Lecomte,
Centre Suisse d'Electronique et de Microtechnique (CH);
D. Calonico, F. Levi,
Istituto Nazionale di Ricerca Metrologica (I);
and members of the SOC2 teams
www.soc2.eu

*Abstract*— The use of ultra-precise optical clocks in space ("master clocks") will allow for a range of new applications in the fields of fundamental physics (tests of Einstein's theory of General Relativity, time and frequency metrology by means of the comparison of distant terrestrial clocks), geophysics (mapping of the gravitational potential of Earth), and astronomy (providing local oscillators for radio ranging and interferometry in space). Within the ELIPS-3 program of ESA, the "Space Optical Clocks" (SOC) project aims to install and to operate an optical lattice clock on the ISS towards the end of this decade, as a natural follow-on to the ACES mission, improving its performance by at least one order of magnitude. The payload is planned to include an optical lattice clock, as well as a frequency comb, a microwave link, and an optical link for comparisons of the ISS clock with ground clocks located in several countries and continents. Undertaking a necessary step towards optical clocks in space, the EU-FP7-SPACE-2010-1 project no. 263500 (SOC2) (2011-2015) aims at two "engineering confidence", accurate transportable lattice optical clock demonstrators having relative frequency instability below $1\times10^{-15}$ at 1 s integration time and relative inaccuracy below $5\times10^{-17}$. This goal performance is about 2 and 1 orders better in instability and inaccuracy, respectively, than today's best transportable clocks. The devices will be based on trapped neutral ytterbium and strontium atoms. One device will be a breadboard. The two systems will be validated in laboratory environments and their performance will be established by comparison with laboratory optical clocks and primary frequency standards. In order to achieve the goals, SOC2 will develop the necessary laser systems - adapted in terms of power, linewidth, frequency stability, long-term reliability, and accuracy. Novel solutions with reduced space, power and mass requirements will be implemented. Some of the laser systems will be developed towards particularly high compactness and robustness levels. Also, the project will validate crucial laser components in relevant environments. In this paper we present the project and the results achieved during the first year.

## I. INTRODUCTION

The principle of an optical lattice clock is shown in Fig. 1. A laser interrogates an ensemble of ultracold atoms, by exciting them to a long-lived atomic state via the clock transition. The atoms are trapped inside a standing-wave laser field ("optical lattice") and possess a temperature of a few micro-Kelvin. The interrogation results in a signal proportional to the absorption of the laser light, which depends on the laser frequency $\nu$. The signal is maximum when the laser frequency coincides with the center of the atomic resonance $\nu_0$. With a feedback control, the laser frequency $\nu$ is continuously kept tuned on $\nu_0$. The resulting ultra-stable laser optical frequency $\nu_0$ can be converted to an equally stable radio-frequency by means of a frequency comb.

The operational procedures in a lattice clock are shown in Fig. 1, middle. An atomic beam produced by an oven travels towards the right through a spatially varying magnetic field. In it, the atoms are slowed down by a laser beam (blue arrow, from laser subsystem BB 1, see Fig. 1 bottom) that finally nearly stops and traps the atoms inside the experimental chamber (square), in the $1^{st}$ stage of a magneto-optical trap (MOT). Subsequently, they are cooled further to a lower temperature of several micro-Kelvin in a $2^{nd}$ MOT stage by another laser subsystem (BB 2). In a third step, they are then transferred to an "optical lattice" made of counterpropagating laser waves generated by a third laser subsystem (BB 4).

Since the lattice potential "wells" are deeper than the thermal energy of the atoms, they are trapped in the potential minima. There, the atoms are spatially localized in one dimension to well below one wavelength of the clock laser (BB 5) and this condition leads to an excitation spectrum free of 1st-order Doppler spectral broadening or shift. Perturbing effects of the lattice light field on the atomic energy levels of the clock transition are minimized by choosing a so-called "magic" wavelength [1, 2].

The clock transition excitation is performed by the clock laser (BB 5). The laser subsystem BB 3 furnishes auxiliary laser light. The subsystems of a lattice clock are shown in Fig. 1, bottom. The laser light produced by the laser breadboards is transported via optical fibers to diagnostic and frequency stabilization units, to the frequency comb, and to the vacuum chamber containing the atoms.

The work of this project involves developing all subsystems, in part in form of compact breadboards, and integrating them into two transportable clocks, one operating with strontium (Sr), one with ytterbium (Yb).

$^{87}$Sr and $^{171}$Yb are currently considered as the most promising isotopes.

TABLE I. SPECIFICATIONS FOR THE LASER AND STABILIZATION SUBSYSTEMS FOR THE STRONTIUM CLOCK

| Laser breadboard | Wavelength | Frequency stability | Size in cm$^3$, mass | Power at fiber outputs |
|---|---|---|---|---|
| BB 1: Cooling #1 | 461 nm ECDL+SHG | 1 MHz | 60×45×10 20 kg | 140 mW to distribution breadboard |
| BB 1: Cooling #1 distribution | 461 nm | 1 MHz | 30×45×10 12 kg | 50 mW MOT, 30 mW slower, 1 mW detection, 1 mW (IR) FSS |
| BB 2: Cooling #2 | 689 nm ECDL | < 1 kHz in 1 h, linewidth < 1 kHz with additional FM | 60×45×12 20 kg | 10 mW MOT, 1 mW FSS 1 mW to stirring breadboard |
| BB 2: Stirring and spin polarisation. | 689 nm | Offset phase locked to 689 nm cooling | 30×45×10 12 kg | 10 mW MOT, 2×1 mW spin polarization |
| BB 3: Repumper | 707 nm, 679 nm ECDL | FM +/- 3 GHz with a few kHz modulation frequency; center: 100 MHz | 30×45×10 12 kg | 2 mW each wavelength to MOT, 1 mW to FSS |
| BB 4: Lattice | 813 nm ECDL | < 10 MHz in 10 h | 30×45×10 12 kg | >200 mW to atomics 1 mW to FSS |
| BB 5: Clock laser | 698 nm ECDL | <1 Hz | 60×45×12 20 kg | 2 mW to atomics, 2 mW to comb, 0.5 mW to cavity, 1 mW to FSS |
| BB 5: Clock cavity | 698 nm | thermal noise 5×10$^{-16}$ | 55×55×55, 30 kg | |
| FSS: Frequency stabilization system | all, exclud., 698 nm | To levels indicated above | 30 × 20 × 10, 10 kg | fiber input for the above fiber outputs |

## II. LASER SYSTEM FOR STRONTIUM CLOCK

The design of the laser system for the Sr lattice clock is modular, and all modules are connected by optical fibers to the vacuum apparatus. This choice ensures high stability and reliability needed for long-term operation of a clock. Moreover, the modular approach allows for independent testing of the subcomponents and, during the course of the project, simple replacement of components by more advanced components. The developments in the field of commercial lasers have produced impressive improvements in size, mass, stability and reliability of lasers and other optoelectronic components. The result is that the complete laser system for the Sr lattice optical clocks is based on off-the-shelf commercial components of moderate size and high reliability, such as diode lasers. The specifications are given in Table I.
Based on ruggedized commercial systems, the lasers (Toptica) have been integrated into compact subsystems where several output beams, controlled in amplitude and frequency via acousto-optical modulators, are produced: a main output to the atomics package, an output for the laser frequency stabilization subsystem, and other additional service outputs. All outputs are provided in single-mode, polarization-maintaining fibers.

The following laser sources have been developed and have been integrated with the atomic package (see Sec. IV): a frequency doubled diode laser (461nm) for 1st cooling (BB 1), a high-power 813 nm laser for the dipole lattice trap (BB 4), two repumper lasers at 679 nm and 707 nm (BB 3). The highly frequency-stable 689 nm laser for 2nd stage cooling on the intercombination transition (BB 2) and the clock laser at 698 nm for the spectroscopy of the clock transition will soon also be integrated.

Fig. 2 shows the Sr laser subsystems developed during the first year of the project. Together, they occupy a volume of ca. 300 liter with an approximate power consumption of 100 W and 150 kg mass.

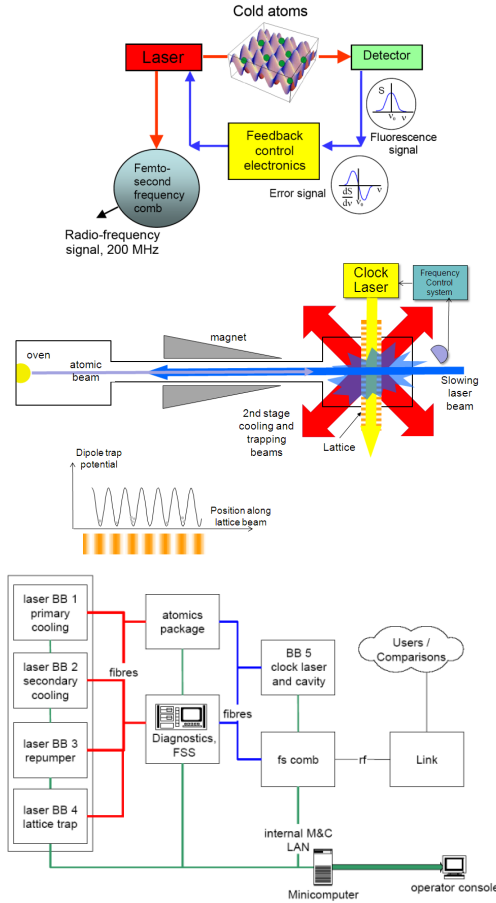

Figure 1. Top left, principle of an optical atomic clock based on atoms trapped by laser light. Top right, schematic of a lattice clock apparatus. Red: laser beams for 2nd stage cooling and trapping in the MOT. Orange-red: lattice laser standing wave; yellow: clock laser wave. Inset: variation of the potential felt by the atoms due to the lattice laser. Bottom, subsystems of a lattice optical clock.

The research leading to these results has received funding from the European Union Seventh Framework Programme ([FP7/2007-2013]) under grant agreement n° 263500.

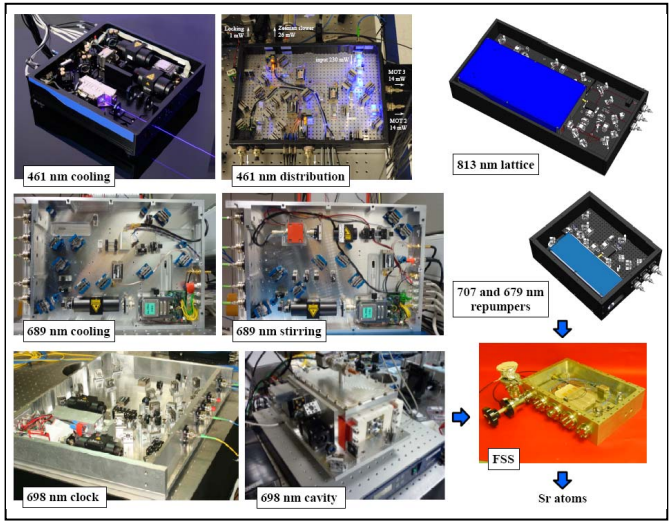

Figure 2. Overview of the laser subsystems for the strontium lattice clock, comprising 7 lasers and the respective frequency stabilization units (698 nm cavity of BB 5 and FSS).

### III. LASER SYSTEM FOR THE YTTERBIUM CLOCK

For the operation of an Yb optical lattice clock, lasers with five different wavelengths are required. In the project, we develop compact and transportable laser sources based on state-of-the-art diode and fiber laser technology, see Fig. 3.

For the 1$^{st}$-stage cooling radiation at 399 nm and for the lattice laser at 759 nm external-cavity-diode-lasers (ECDL) based on narrow-band interference filters [3, 4] are being developed, which promise improved stability compared to the commonly used grating-stabilized ECDLs. A prototype interference-filter ECDL at 399 nm using standard laser diodes and delivering up to 40 mW has already been tested successfully as a master laser. It is used to inject another, free-running laser diode and provides light for initial atom slowing. The higher power of a few 100 mW that is required at the optical lattice wavelength of 759 nm will be achieved using a self-injected tapered amplifier.

A compact repumping laser at 1389 nm, required to reduce fluctuations of the clock interrogation signal, has been developed following the approach at NIST. The unit is based on a DFB laser diode with fiber output, and exhibits a low free-running frequency instability (about 15 MHz/day linear drift) that will allow using the laser without further frequency stabilization.

The postcooling laser at 556 nm is a laser system based on fiber laser technology. It is designed to have a total volume of 3 liter. The all-fiber optical setup consists of three stages. The seed signal at 1111.60 nm is generated by a NKT Photonics BASIK module. The infrared signal is amplified in an amplifier pumped by two pump laser diodes at 974/980 nm. Second harmonic generation (SHG) at 555.80 nm is performed by an all-fiber coupled waveguide periodically poled lithium niobate (PPLN) device. An output power of 20 mW has been achieved.

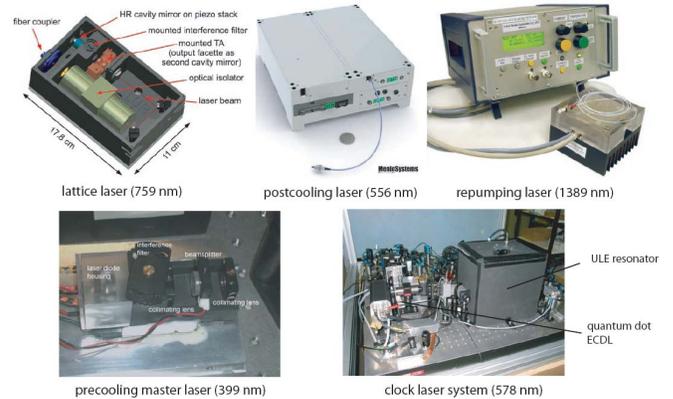

Figure 3. Overview of the laser subsystems developed for the ytterbium lattice clock, comprising 5 lasers.

Our approach for providing the 578 nm clock radiation is based on SHG of the radiation of an external cavity quantum dot laser (QD-ECDL) at 1156 nm in a PPLN waveguide. Stability and linewidth at the 1 Hz level have been achieved by stabilizing the laser to a highly stable ULE reference cavity [5].

### IV. STRONTIUM ATOMS PREPARATION AND PRELIMINARY CLOCK SPECTROSCOPY

The design of the first-generation compact vacuum apparatus [6, 7] and the clock breadboard, which is fully operational, are shown in Fig. 4. Strontium atoms are first evaporated by an efficient oven working at 420°C, with a power consumption of 34 W. The atomic beam is then decelerated in a 18 cm long Zeeman slower and finally loaded into a 1$^{st}$-stage MOT, operating on the dipole-allowed $^1S_0$ - $^1P_1$ transition at 461 nm. Radiation at this wavelength and for repumping at 679 nm and 707 nm are produced by the laser systems described above. Initial work has been performed with $^{88}$Sr. The typical number of loaded atoms is about $10^8$. By observing the expansion of the atoms from the MOT, an atomic temperature of about 2 mK was determined.

The 2$^{nd}$-stage MOT operates on the $^1S_0$ - $^3P_1$ transition at 689 nm. Pending availability of the newly developed 689 nm laser, a prototype master-slave laser delivering up to 50 mW has been employed. The slave laser is optically injected with a beam coming from the pre-stabilized master laser. In order to tune the laser frequency on resonance and to provide the necessary power level and frequency modulation for the 2$^{nd}$-stage MOT, a double-pass acousto-optical modulator (AOM) is used.

As shown in Fig. 5, in the 2$^{nd}$-stage MOT two phases are implemented, a 120 ms long "broadband" phase during which the frequency of the cooling laser is broadened to 5 MHz to cover the Doppler width of the atomic resonance of the atoms at the end of the 1$^{st}$-stage MOT, followed by 30 ms of "single-frequency" phase (no broadening). With the broadband phase it is possible to cool and trap about 1 x $10^7$ $^{88}$Sr atoms at 22 μK, while the "single-frequency phase" further cools the atomic sample to the 2 µK, with a final population of 1 x 10$^6$ atoms.

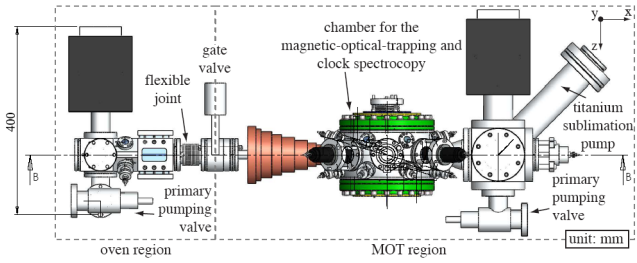

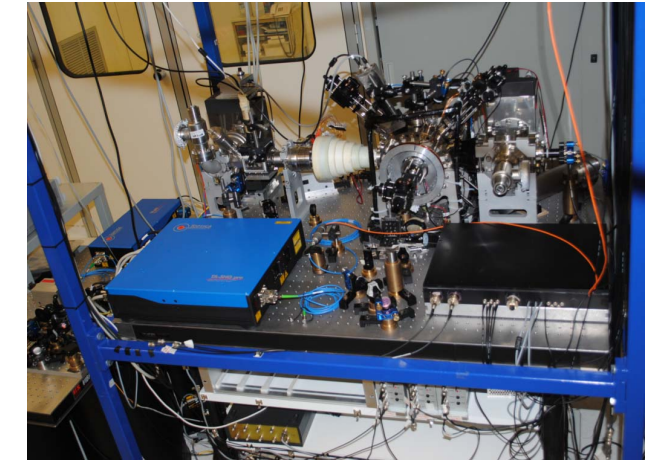

Figure 4. Top: Section view of the 1$^{st}$ generation strontium atomics package. The extension of the vacuum system is ca. 110 cm x 35 cm x 40 cm (150 liters). Bottom: clock breadboard with some of its laser subsystems.

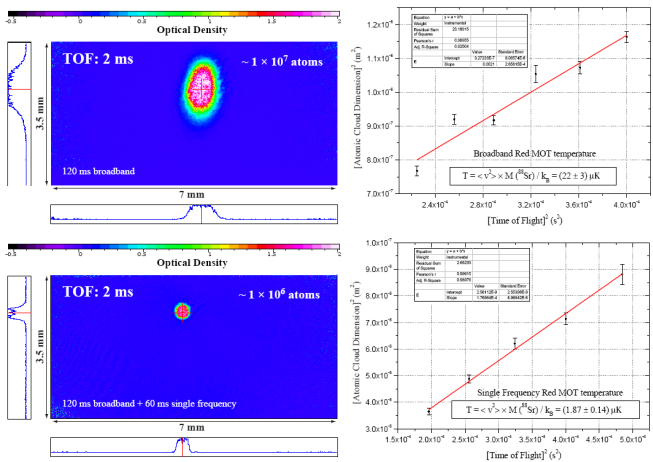

Figure 5. Absorption images of the $^{88}$Sr sample and time-of-flight (TOF) measurements of the atom cloud size (graphs) at the end of the "broadband" (top) and "single-frequency" (bottom) phase.

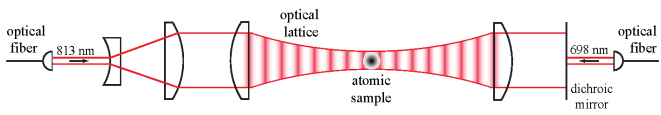

Figure 6. Optical setup of the vertical optical lattice at 813 nm, showing also the clock radiation input.

Subsequently, the $^{88}$Sr atoms are then loaded into a vertical 1D optical lattice, realized by means of the external-cavity diode laser at 813 nm developed for this project. In Fig. 6 the design of the optical setup implemented for the vertical 1D lattice is shown. Light coming from the 813 nm laser source is sent through a beam expander, focused on the atomic cloud at the center of the main chamber, recollimated and then retro-reflected. With an available power of 280 mW and a waist of 50 µm, the estimated lattice trap depth is about 5 µK. The infrared beam has been aligned on the cold atomic cloud with the help of a resonant blue beam copropagating with the infrared beam.

Fig. 7 shows an absorption image of the atomic sample trapped in the optical lattice after the single-frequency 2$^{nd}$ stage MOT phase. About 50% of the atoms ($5\times10^5$) are transferred from the latter into the lattice trap. The observed lifetime of the atoms in the trap is about 1.4 s, indicating that heating effects due to amplitude or frequency noise coming from the (unstabilized) 813 nm source are low.

For a preliminary clock spectroscopy test, the stationary 698 nm clock laser developed in Firenze [8] was used, which has a frequency stability of 10$^{-15}$ for integration times between 10 - 100 s. Laser light, resonant with the $^1S_0$-$^3P_0$ clock transition, is coupled through a dichroic mirror along the direction of the lattice (Fig. 6). An AOM is employed to precisely control the timing, frequency and intensity of the excitation pulse on the atomic cloud. Preliminary results of magnetic-field-induced spectroscopy on the clock transition for the $^{88}$Sr isotope are shown in Fig. 8. For these measurements we applied a constant magnetic field $B$ (along the polarization of the clock laser field and the trapping field) by inverting the current on one of the MOT coils. By reducing the magnetic field ($B$ = 1.1 mT) and the interaction time ($T$ = 300 ms) we observed a minimum linewidth of about 410 Hz (left plot in Fig. 8).

As demonstrated in [9] it is possible to find the transition on a day-to-day basis even without a precise calibration of the laser frequency by adding a 200 kHz chirping (with 2 s period) on the clock laser frequency and increasing the interaction time to 1 s.

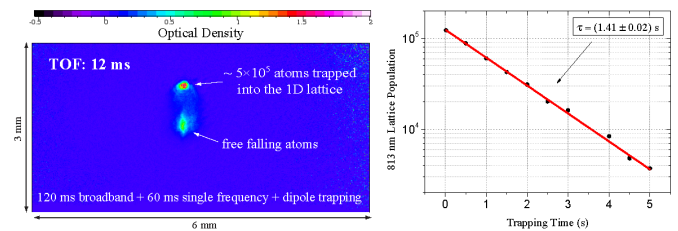

Figure 7. Left: absorption image taken 12 ms after turning off the 689 nm beams, while the 1D lattice (813 nm) is applied. A significant fraction of the atoms remain trapped in the lattice, while the untrapped fraction falls in the gravitational field. Right: Determination of the lifetime of $^{88}$Sr atoms trapped in the lattice via measurement of the number of atoms remaining trapped for different durations of the applied optical lattice.

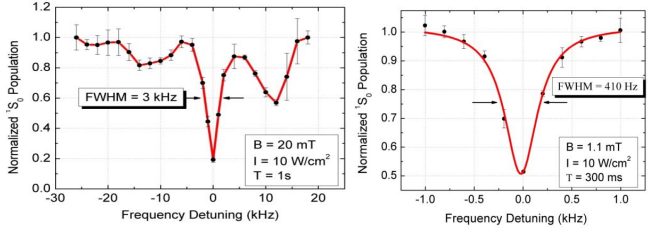

Figure 8. Spectroscopy of the $^{88}$Sr clock transition in the optical lattice. The sidebands seen in the left plot indicate confinement of atoms in the lattice.

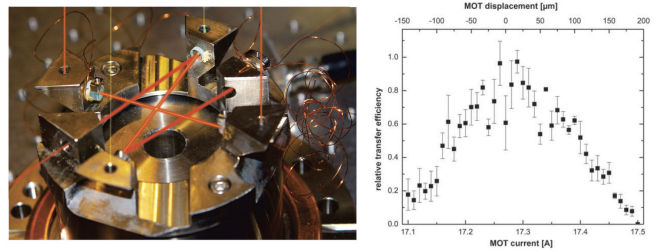

Figure 9. Left, the monolithic resonator setup for 1D and 2D optical lattices. Right, transfer efficiency into the optical lattice as a function of 2$^{nd}$ stage MOT position. The MOT position is controlled via the current through one of the MOT magnetic field coils.

## V. ATOM PREPARATION IN THE YTTERBIUM CLOCK

Within the first year of the project we have succeeded in preparing ultracold ensembles of the two isotopes $^{174}$Yb and $^{171}$Yb in a one-dimensional optical lattice which operates at 759 nm, the magic wavelength for Yb [10]. Our approach for the realization of such a magic-wavelength optical lattice includes enhancement resonators which are placed in the vacuum chamber in which the cold Yb ensembles are prepared. Inside two perpendicular resonators a large-volume optical lattice (either one- or two-dimensional) can be formed with only a few 100 mW power from a diode laser (Fig. 9 left). The resonator mirrors, two of which are mounted on ring piezo actuators, are glued to a monolithic structure, made out of the steel alloy invar, in order to increase the passive stability of the optical lattice. The end mirrors are transparent for the clock transition wavelength of 578 nm, which makes it simple to superimpose the radiation at the clock wavelength with the optical lattice.

Loading of a 1D optical lattice was so far achieved by using one of the two enhancement resonators with a beam waist of ca. 150 μm. We estimate that currently the maximum achievable trap depth is on the order of 50 μK. Yb atoms are loaded into the lattice by carefully aligning the position of the 2$^{nd}$ stage MOT to the lattice position and ramping down the 556 nm cooling light field (see Fig. 9, right). Successful loading of the optical lattice is observed by turning the cooling light field back on after a variable hold time and detecting the fluorescence of the atom that remained trapped in the lattice in the mean time. Without the lattice light field, no atoms are recaptured after roughly 20 ms while with the light field recapturing is possible even after 300 ms. The longest lifetime in the optical lattice observed so far is 130 ms, sufficient for the operation of an optical lattice clock.

In the currently used prototype setup we can transfer more than 20% of the atoms from the 2$^{nd}$-stage MOT into the optical lattice, amounting to roughly 10$^5$ atoms, as aimed for. Since the temperature of the atoms in the MOT is typically 30-50 μK, we may infer that the transfer efficiency is limited by the depth of the optical lattice. This limit should be overcome in an advanced resonator setup, which is designed to allow for lattice depths of several 100 μK and higher transfer efficiency.

## VI. NEXT-GENERATION SUBUNITS

### A. Blue light generation

The 461 nm cooling light for Sr is obtained by second-harmonic generation. As a potentially simpler and more robust alternative to the conventional generation in an external enhancement cavity (used in the laser developed for this project), we have tested the single-pass generation in periodically poled KTP waveguides. We could couple fibers with waveguides such that the input coupling efficiency to the waveguide was up to 70% (at 922 nm) and could obtain up to 40 mW of power at 461 nm. The output power could not be increased to more than 100 mW, as required. Therefore, an evaluation of different waveguide types is planned to follow.

### B. Vacuum chambers

For a transportable clock, robustness, compactness and moderate mass are desirable. In this respect, we have compared two different techniques for realizing the UHV environment in the atomic package: lead sealing and gluing. Although both techniques have yielded vacuum in the range of 10$^{-11}$-10$^{-12}$ mbar, the gluing technique has resulted in a more compact and lightweight vacuum chamber which can be baked well above 200 °C. In the process, we have analysed different combinations of materials for the chamber, windows and glues. We have found that although several combinations are possible, MACOR/titanium for the chamber material together with BK7/YAG for windows is a good combination from the thermal expansion point of view. We have also found that the two adhesives H77 and 353ND work quite well.

### C. Atom preparation

Two novel approaches for loading atoms into a lattice optical clock are being investigated.

For the first approach, a two-dimensional (2D) MOT loaded from a dispenser is being tested. We have obtained some preliminary results where a 3D MOT is loaded from the 2D MOT system. We have observed clear enhancement in the atom number in the 3D MOT when loaded from the 2D MOT (Fig. 10 inset). Having gained from the above experience, we have made a preliminary design for a 2$^{nd}$ generation compact and lightweight vacuum chamber as required for the project (Fig. 10 main). The design is flexible in the sense that either a 2D MOT or a Zeeman slower can be used for slowing. The ultra-high vacuum will be maintained by ion pumps. In addition to the optical access for lasers for 2$^{nd}$ stage cooling, detection, lattice trapping etc, a thermal enclosure will be designed for the atomics package, which allows control of temperature to better than 0.1 K at the position of the atoms, necessary for controlling the black-body systematic frequency shift. Closed-loop magnetic field control will also be installed.

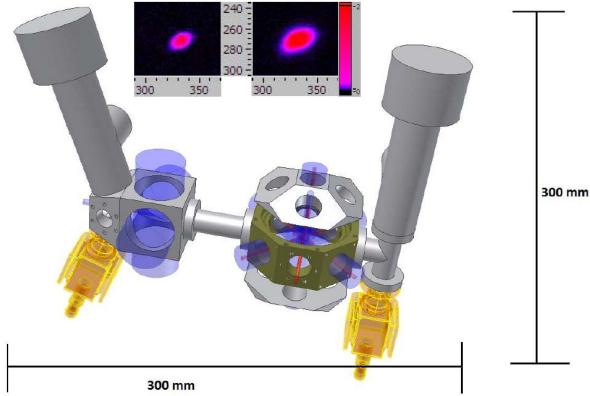

Figure 10. Main figure, preliminary design of the second-generation atomics package, where a 3D MOT (center) will be loaded from a 2D MOT (left side). The design is also adaptable to a Zeeman slower replacing the 2D MOT. Blue: cooling and detection beams; red: 2$^{nd}$ stage cooling and lattice trapping beams; yellow: ion pumps. Inset, two images of a $^{88}$Sr 3D MOT loaded without (left) and with (right) a 2D MOT on, in a separate setup. The enhancement of the atom number is clearly visible.

The second, conventional, approach for capturing atoms and loading them into a lattice trap consists in slowing the hot atoms emitted from the oven using a Zeeman slower, as is done in the 1$^{st}$ generation breadboard (Fig. 4). Typically, in a Zeeman slower a tapered solenoid is used. We have developed a novel transverse magnetic field Zeeman slower [11], which uses permanent magnets situated at adjustable distances from the beam axis (Fig. 11 top), and which works equally well for both $^{87}$Sr and $^{88}$Sr isotopes. This has potential for achieving a clock apparatus with smaller footprint, reduced mass and no dissipation.

### D. Black-body radiation control

Finally, one of the major frequency shifts encountered in neutral atom lattice clocks is the black-body radiation shift due to the surrounding apparatus. Its value needs to be known in order to reach the goal accuracy of the space clock. We have developed a blackbody chamber to test calculations of the Sr black-body shift coefficient. The chamber includes two narrow copper tubes that approximate to well-defined black-body sources (Fig. 11 bottom). Cold Sr atoms at micro-Kelvin temperatures will be transported into either tube by a moving-lattice beam, then transferred into the "magic-wavelength" lattice for interrogation by the clock laser. By sequentially probing the lattice-trapped atoms in the two tubes at different temperatures, the differential blackbody shift is measured, allowing validation of the black-body coefficient.

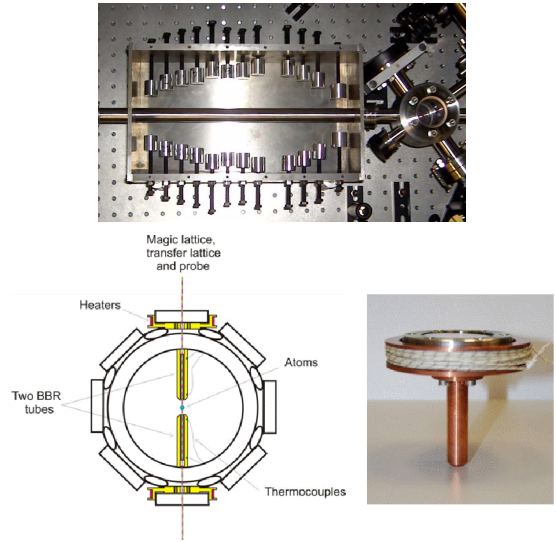

Figure 11. Top, a Zeeman slower using permanent magnets. Bottom left, schematic of MOT/lattice chamber with blackbody tubes. Bottom right, photo of a blackbody tube showing heater section external to the UHV chamber.

### VII. CONCLUSION

The newly constructed Strontium laser subsystems have in part been integrated with the atomics subsystem and have already allowed $^{88}$Sr atoms to be efficiently cooled, trapped into a 1D optical lattice and interrogated on the clock transition. The subunits of the Yb clock are also working well individually, with the atoms routinely trappable in the optical lattice. The upcoming work for the 2$^{nd}$ year of the project will include: (i) completion of the integration, spectroscopy of the clock transition, initial characterizations of the clocks' performances, and optimization; and (ii) progress on the development of the second-generation subunits (lasers, atomics package components), so that they can be integrated into the clocks in year 3.

Preparatory activities are also in progress for the later robustness testing of lasers and for the full characterization of the transportable optical clocks after moving them from the integration labs to national metrology labs, and for next-generation compact optical frequency combs.

At the end of the project we expect to have an operational lattice clock that will represent the baseline design for the production of the flight model for the ISS space clock.

### ACKNOWLEDGMENT

We thank L. Cacciapuoti and O. Minster (ESA) for their continued support.